\documentclass[usenatbib,usegraphicx,twocolumn]{mn2e}
\def\mnras{MNRAS}
\def\mut{\tilde{\mu}}
\def\p{p}
\def\z{{\bf z}}
\def\r{{\bf r}}
\def\x{{\bf x}}
\def\apj{Ap.J}
\def\aj{AJ}
\def\r{{\bf r}}
\def\w{w}
\def\dt{\tilde{\eta}_{\rm HI}}

\def\th{\vec{\theta}}
\def\thpv{\vec{\theta_{\perp}}}
\def\thp{\theta_{\perp}}
\def\thl{\theta_{\parallel}}
\def\x{{\bf x}}

\def\HI{{\rm HI}}
\def\H{{\rm H}}
\def\m{{\bf m}}
\def\mK{\rm mK}
\def\d{\eta_{\HI}}

\def\n{{\bf n}}
\def\Tc{T_\gamma}
\def\Tb{T_b}
\def\Ts{T_s}
\def\bTs{\bar{T}_s}
\def\Tg{T_g}

\def\A10{A_{10}}

\def\k{{\bf k}}

\usepackage{epsfig}
\usepackage{graphics}
\begin{document}
  \title[Anisotropies in redshifted 21 cm maps tell us?]{What will
    anisotropies in the clustering pattern in redshifted 21 cm maps
  tell us?}  
  \author[ S. S. Ali, S. Bharadwaj and B. Pandey]{Sk. Saiyad
  Ali\thanks{Email:     saiyad@cts.iitkgp.ernet.in}, 
  Somnath  Bharadwaj\thanks{Email: somnathb@iitkgp.ac.in} and 
Biswajit Pandey\thanks{Email:  pandey@cts.iitkgp.ernet.in},\\ 
\\ Department of Physics and Meteorology\\ and
\\ Centre for Theoretical Studies \\ IIT Kharagpur \\ Pin: 721 302 ,
  India }
\maketitle
\begin{abstract} 
 The clustering pattern in  high redshift HI  maps   
is expected to be anisotropic due to two distinct reasons, the
 Alcock-Paczynski effect and the peculiar velocities, both of which
 are sensitive to the cosmological parameters. The signal is also
 expected to be sensitive to the details of the HI distribution at the
 epoch when the radiation originated. We use simple models
 for the HI distribution at the  epoch of reionizaation and the
 post-reionization era to investigate exactly what we hope to learn from
 future observations of the anisotropy pattern in HI maps. We find
 that such observations  will probably tell us
 more about the HI distribution  than about the background
 cosmological  model. Assuming that reionization can be described
by  spherical, ionized bubbles all of the same size with  their
 centers  possibly being biased with respect to the dark matter, 
we find that the anisotropy pattern at small angles 
 is   expected to have a bump at the characteristic angular
size  of the individual bubbles   whereas the large scale
anisotropy pattern will  reflect the size and the bias of the
bubbles. The anisotropy also depends on the background cosmological
parameters, but the dependence is much weaker. Under the assumption
that the HI in the post-reionization era traces the dark matter with a
possible bias, we find that changing the bias and changing the
background cosmology has similar effects on the anisotropy
pattern. Combining observations of the anisotropy with independent
estimates of the bias, possibly from the bi-spectrum, may allow these
observations to constrain cosmological parameters.

\end{abstract}
\section{Introduction}
Observations of redshifted 21 cm radiation from the large-scale
distribution of neutral hydrogen (HI)  are perceived 
one of the most important future probes of  the universe at high 
redshifts, and  there currently are   several initiatives underway
towards carrying out such observations.  To list a few, the GMRT
\citep{swarup} is already functioning at several bands in the
frequency range $150$ to $1420  \, {\rm MHz}$,  while 
LOFAR\footnote{http://www.lofar.org }, 
PST\footnote{http://astrophysics.phys.cmu.edu/$\sim$jbp } and 
SKA\footnote{http://www.skatelescope.org} are in different stages of
design and/or construction.  
These observations hold the potential of  probing  the HI evolution  
through epochs  starting  from  the  present  all the way to  $z 
\sim 50$ when the universe was in the  Dark Ages \citep{miralda}. Variations with 
angle and with frequency (or redshift) of  the redshifted 21 cm
radiation brightness temperature give  three  dimensional maps  of
the HI distribution at high $z$ \citep{hogan}. The signal 
is  sensitive to the   spin temperature $\Ts$ of
the HI hyperfine 
transition,  and    at any $z$ the HI will be observed in
emission or absorption depending on whether  $\Ts$ is larger or
smaller than $\Tc$ the CMBR temperature. To get a picture of the 
expected signal, we outline below how  $\Ts$ evolves with $z$. 

The evolution of $\Ts$ is governed by two tendencies, once which
drives it towards $\Tc$ and another towards $\Tg$ the  HI gas
kinetic temperature.  $\Ts$  couples to  $\Tc$ through a radiative
transition,  and  it couples to $\Tg$ through atomic
collisions and   through the absorption and re-emission of
Ly-$\alpha$ photons (the 
Wouthuysen -Field effect, \citealt{uthu},\citealt{fild} ).   
The collisional process dominates 
at high densities where it causes $T_s$ to closely follow 
$\Tg$. The small density of free electrons, residual  after hydrogen 
recombination at $z\sim 1000$, transfers energy from the
CMBR to the gas and maintains  $\Ts=\Tg=\Tc$ up to $z \sim 200$. The
heat transfer is  ineffective at $z < 200$,  and the HI  cools
adiabatically whereby $\Ts=\Tg<\Tc$. The collisional process which
maintains $\Ts=\Tg$ looses out to the radiative process at $z < 30$
and $\Ts$ approaches $\Tc$.  Thus, there is a redshift range $200 \ge
z \ge 30$ where $\Ts < \Tc$ and the HI will be seen in absorption
against the CMBR \citep{scott}. The HI at these epochs largely traces
the  dark matter, and  the  HI maps  would probe the power-spectrum of 
density perturbations at a high level of precision (\citealt{lz};
\citealt{bharad5}). The first luminous objects, formed 
at $z \sim 30$,   would heat the gas through the emission of 
soft X-ray photons (\citealt{chen}; \citealt{rikoti} )and through
Ly-$\alpha$ photons which would also couple 
$\Ts$ to $\Tg$ through the  Wouthuysen -Field effect. The HI gas 
will now be partially heated , and the coupling of $\Ts$ to $\Tg$ will
also be partial depending on the local flux of Ly-$\alpha$ photons.
This is an additional source of fluctuation in the HI  maps,  and it
may be possible to  detect the presence of the first 
luminous objects in the redshift range $30 \le z \le 15$ through this
effect (\citealt{barkana1}, 2004b). The coupling of $\Ts$ to  
$\Tg$ is expected to be saturated by  $z \sim 15$,
and $\Ts=\Tg \gg \Tc$ throughout the gas  {\it ie.} the  HI will be
seen in emission and the HI maps will again trace the  dark
matter. This changes at $z \sim 10$ 
when a substantial fraction of the HI is ionized by the first luminous
objects. During the epoch of reionization, the HI maps  trace the
size, shape and topology of the ionized regions
(\citealt{gnedin}; \citealt{ciardi}; \citealt{sokasiana}, 2003b;
 \citealt{furlanetto1}, 2004b). At low redshifts ($z<6$), the 
bulk of the HI is in the high column density clouds seen as DLAs in
quasar spectra (\citealt{peroux}; \citealt{lombardi};
\citealt{lanzetta}). These clouds are possibly disk  galaxies or
their progenitors, and they trace  the dark matter,  may be with some
bias. The  HI will be seen in emission, and the maps   will trace the
dark matter   (\citealt{bharad1}; \citealt{bharad4}). 

Though the HI signal in each redshift range will probe a different phase
 of the  HI evolution and have it's own distinct signature, there is 
one thing in common throughout in that the clustering pattern will be
 anisotropic. 
 The anisotropies in the clustering pattern are the results of two
 distinct effects. The first is the Alcock-Paczynski
 effect (\citealt{alcock}, hereafter the AP effect) caused by the non
 Euclidean geometry of space time and the second is the  effect of 
 peculiar velocities. 
The  AP effect causes  objects which  are 
 intrinsically spherical in real space to appear  
elongated along the line of sight in redshift space. This effect is  
particularly important at high redshifts  ($\rm z \ge .1$) where it is 
sensitive to the cosmological parameters which it can be used to probe. 
The two-point correlation function of the gravitational  clustering of
 different kinds of objects is expected
to be statistically isotropic, and this is a natural choice for applying 
this test. Galaxy redshift surveys do not extend to sufficiently high 
redshifts for this effect  to be significant. Quasar redshift surveys
 extend to much higher redshifts and there has been a considerable
 amount of work (eg.  \citealt{balli} , \citealt{matsu} ,
 \citealt{naka}  , \citealt{popo} ,\citealt{nair}) investigating the
 possibility of determining the parameters $\Omega_m$ and  
$\Omega_\Lambda$. The main problem is that quasar surveys
 (eg. SDSS) are very sparse and hence they are not optimal for
 determining  the cosmological parameters
 (\citealt{matsu1}). \citet{hui} have used the AP test to constrain  
 cosmological parameters using the Lyman-alpha forest.

The redshift, used to infer  radial distances,  has a contribution
from the line of sight component of the peculiar velocity and this
introduces a preferred direction in the redshift space clustering
pattern. There are two characteristic effects of peculiar
velocities.  On small scales, the random motions in  virialized
regions causes the clustering  pattern  to appear elongated along the
line of sight in redshift space.  
On large scales the coherent infall onto clusters and superclusters,   
and the outflow from  voids causes the redshift space 
clustering pattern to appear  compressed along
the line of sight (the Kaiser effect, \citealt{kais}).  This  effect
can be modeled using 
linear theory  and the 
anisotropies in the redshift space clustering pattern can be used to
determine the parameter $\beta=\Omega_m^{0.6}/b$ where $\Omega_m$  and
$b$ 
are the cosmic mass density parameter and  the linear bias parameter
respectively (\citealt{kais}, \citealt{ham}). 
The Kaiser effect has been studied in the different  large galaxy redshift
surveys (eg. SDSS , 2dfGRS) and a recent investigation of the redshift
space distortions in the 2dFGRS yields a value of $\beta=0.49\pm 0.09$
(\citealt{haw})  at an effective redshift
$z\approx 0.15 $ .

Future  redshift survey carried out using redshifted 21 cm HI   radiation 
will allow the anisotropies arising from both the AP effect and the
peculiar velocities to be studied to redshifts as high as $z \sim 50$
and possibly higher, surpassing the potential of any future galaxy or
quasar survey. 
In a recent paper \citet{adi} has studied  the AP effect  in
redshifted 21 cm maps of  the epoch of reionization  and  he has
proposed that a determination 
of the correlation of temperature fluctuations to an accuracy of $20
\%$ should allow a successful application of the AP test to constrain
the background cosmological model.

\citet{barkana1}  have proposed that measuring the anisotropy of the
HI power spectrum arising from peculiar velocities provides a means
for distinguishing between the different sources which contribute to
HI fluctuations. Further, they propose that it may be possible to
separate the   primordial inflationary power
spectrum  imprinted in the dark matter from the various astrophysical
sources which will also contribute to the HI fluctuations.

In this paper we re-examine exactly what  we hope to learn from
observations of the anisotropy in HI maps. The previous analysis of 
\citet{barkana1} using the power spectrum implicitly assumes that the 
 background cosmological model is known to a great level of precision
 and hence does not take into account the anisotropies introduced by
 the geometry (AP effect).  We adopt a framework
which allows high redshift HI observations to be interpreted  without
reference to a background cosmological model. We use the HI
temperature two-point correlation function  which deals
with  directly observeable quantities. 
 Using this framework, we  
quantify the anisotropy and study its dependence on the background
cosmological model. The AP effect and the Kaiser effect differ in
their response to variations in the background cosmological model. 
To estimate the relative contributions from these two effect we  also
calculate the anisotropy ignoring the peculiar velocities. 

Our work focuses on the  signal expected from
the epoch of reionization and the post re-ionization era, these being
the most promising frequency bands for observations in the near
future. 
 The HI distribution in 
these eras is largely unknown, and it is expected to differ  
from  the dark matter distribution. The HI  is expected to have a
very patchy distribution at epoch of reionization. Determining the size,
shape and distribution of the 
ionized regions is one of the main forces driving the effort towards
future HI observations. The post-reionization HI will probably 
trace the dark matter with a bias. In addition  the parameters of
the background cosmological model, the anisotropy in the HI maps is
expected to also be  sensitive to the details of the high redshift HI
distribution, the latter being largely unknown. In this paper we use
simple models for the HI distribution to ask if observations of the
anisotropy tell us more about the background cosmological models or
the details of the HI fluctuations.

This paper is organized as follow.  In Section 2. we explain the
origin of the anisotropies and how they can be quantified. 
 In Section 3. we present the background cosmological models and
models for the HI  distribution.  Section 4 contains the results and
in Section 5. we present discussion and conclusions. 
 
It may be noted that unless mentioned otherwise, we use the values 
$ (\Omega_{b} h^2,h)=(0.02,0.7)$
throughout. 
\section{Origin of the anisotropies}
Observations of the HI radiation  
at a  redshift $z$  along the direction of the unit vector $\m$
will measure 
\begin{equation}
\delta T_b(\m,z)=\frac{T_b(\m,z)-\Tc}{1+z}\, ,
\end{equation}
 $T_b(z,\m)$  being the brightness temperature of the
 HI radiation at the position and epoch when the radiation originated
 and $\Tc$  the temperature  of the CMBR also at the same epoch.  It
 is convenient to represent such observations as 
$\delta \Tb({\bf z})$ where   ${\bf   z}= z \, \m$ denotes  the
space of possible redshifts and directions of observation which  we refer
to as  redshift space.  Here we assume that the redshift range and
region of the sky under observation are both small. Under this
assumption, the separation between any two points in redshift space is 
\begin{equation}
{\bf \Delta  z}= \Delta z \, \n + z \thpv  
\label{eq:f1}
\end{equation}
where $\n$ is the line of sight to the center of the region being
observed, $z$ is the mean redshift, $\Delta z$ the difference in
redshift and $\th$ is the angular separation which is a  two
dimensional vector in the plane of the sky. It may be noted that 
$\Delta z \,\n$ and $z \thpv$ are respectively the components of ${\bf
  \Delta z}$   parallel and perpendicular to the line of sight $\n$,
  and we introduce the notation  $\thl=\delta z/z$ whereby 
\begin{equation}
{\bf \Delta  z}= z \, (\thl \, \n +  \thpv)  
\label{eq:f2}
\end{equation}

\begin{figure}
   \includegraphics[width=84mm]{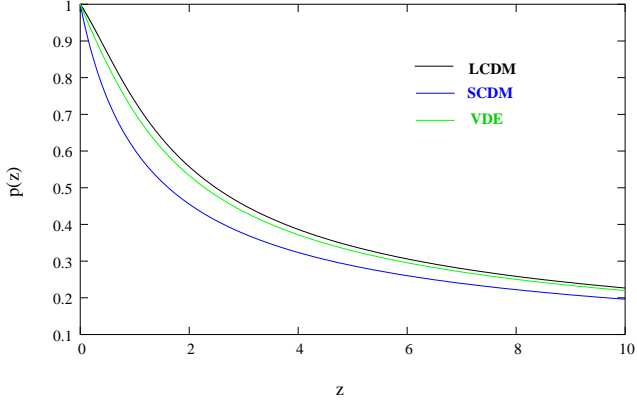}
  \caption{ This  shows the redshift dependence of the   geometrical
  distortion   parameter $p(z)$ for  different background
  cosmological models. } 
  \label{fig:a1}
 \end{figure}

We next shift our attention to quantifying the fluctuations in the HI 
  brightness temperature.  The temperature two point correlation
  function  
\begin{equation}
 \w(\thl,\thp)=\langle \delta \Tb ({\bf z}) \,  \delta \Tb (
{\bf z +  \Delta z})  \rangle
\label{eq:f3}
\end{equation}
is the statistical quantity of our choice.  Here we have invoked the
property of  statistical isotropy whereby $\w$ is independent of the
direction of $\thpv$.   The point to note is that  $\w(\thl,\thp)$
characterizes the observations  solely in terms  of directly
observable quantities, namely angles and redshifts.  and does not
refer to a background cosmological model. It is necessary to assume a
cosmological model in order to assign a physical position  ${\bf  r}$
to a vector  ${\bf  z}$ in redshift space. Using $r(z)$ to
denote the comoving distance corresponding to a 
redshift $z$, 
\begin{equation}
r(z)= \int^{0}_{z}\frac{c \, dz^{'}}{ H(z^{'}) }
 \label{eq:a1}
\end{equation}  
where $H(z)$ is the Hubble parameter, 
we can assign the vector ${\bf r}= r(z) \, \m$ in real
space to the vector ${\bf z}= z \, \m$ in redshift space.  
The physical separation ${\bf \Delta r}$ (in comoving
coordinates) corresponding to the separation ${\bf \Delta z}$ in
eq. (\ref{eq:f2}) is 
\begin{equation}
{\bf \Delta  r}= r(z) \, \left[ \p(z)  \, \thl \,  \n +  \thpv) \right] 
\label{eq:f4}
\end{equation}
where $\p(z)=d  \ln  r(z)/d \ln  z$. The presence of the term $\p(z)$
in eq. (\ref{eq:f4})  makes the mapping from redshift space to real
space anisotropic  when $\p(z) \neq 1$. At low redshifts  $(z \ll 1)$
we have  $r=cz/H_0$ whereby  $\p(z)=1$ and ${\bf \Delta  r}= (c/H_0)
\, {\bf \Delta  z}$ {\it ie. } the mapping from redshift to real space
is   isotropic. The high redshift behaviour of $\p(z)$  depends on 
the cosmological model (Figure \ref{fig:a1}) with the feature
$\p(z)<1$ being common to all the models. This introduces an
anisotropy, and a sphere ${(\bf \Delta r})^2=R^2$ in real space will
appear as an ellipsoid $(\thl \, \p)^2 + (\thpv)^2=(R/r)^2$
elongated along  the line of sight in redshift space.
This is the Alcock-Paczynski \citep{alcock}  effect. For a
fixed redshift,  
the elongation depends on the cosmological model. For all models, the
elongation increases with $z$. 

We next discuss how the AP effect will manifest itself in
$\w(\thl,\thp)$. Assuming for the
time being that the fluctuations in the HI brightness temperature
directly trace the fluctuations in  the HI density at the point where
the radiation originated {\it ie.} $\delta T_b({\bf z}) = K \,
\delta_{\rm   HI}({\bf r})$ we have 
\begin{equation}
 \w(\thl,\thp)= K^2 \,  \xi_{\rm HI} ( {\bf \Delta r},\,z)
\label{eq:f5}
\end{equation}
where $\delta_{\HI}$ denotes  fluctuations in the HI density,
$\xi_{\HI}( {\bf \Delta r})=\langle \delta_{\HI}({\bf r})
\delta_{\HI}({\bf r+ \Delta r}) \rangle $
is the two-point correlation function of $\delta_{\HI}$ and $K$ is a
 proportionality factor. We expect the fluctuations in the HI density
 to be statistically isotropic {\it ie.}  $\xi_{\HI}$ will  not depend on
 the direction of ${\bf \Delta r}$. It then follows from the
 preceding discussion that $\w(\thl,\thp)$ will be anisotropic when
 $\p(z) \neq 1$, and  the curves of constant $\w(\thl,\thp)$ will be
 ellipses elongated along the line of sight.

 The analysis till now has ignored the effect of peculiar
velocities. The line of sight component of the peculiar velocity makes
an additional contribution to the redshift, and  eq. (\ref{eq:a1})
does not  correctly predict the   comoving distance. On large scales
the peculiar velocity field is well described using linear perturbation
theory which relates it to  density fluctuations in the underlying dark
matter distribution.  The coherent flows into over-dense
regions and out of under-dense regions leads to the compression of
structures along the line of sight \citep{kais}. The introduces an
an anisotropy pattern which is opposite in nature to the AP effect. 

\begin{figure}
  \includegraphics[width=84mm]{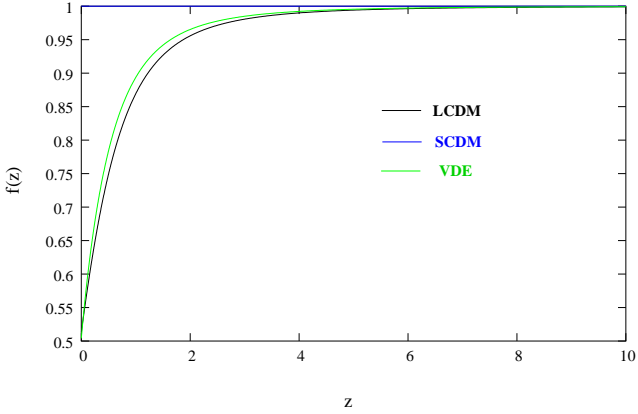}
  \caption{This shows the redshift dependence of the  peculiar
    velocity distortion 
    parameter $f(z)$ for  different
    background cosmological models. It may be noted that for the SCDM
    model 
    $f(z)=1$ throughout.} 
  \label{fig:a2}
\end{figure}

Incorporating the effect of the coherent flows to linear order, the
fluctuations is the brightness temperature of the HI radiation has
been calculated  (\citealt{bharad5},\citealt{bharad6}) to be
\begin{equation}
\delta \Tb(\z) = \bar{T}(z) \times \d(\r,z)
\end{equation}
where  
\begin{equation}
  \bar{T}(z)=4.0 \, \mK   (1+z)^2  \, \left(\frac{\Omega_b
    h^2}{0.02}\right)  \left(\frac{0.7}{h} \right) \frac{H_0}{H(z)}   
  \label{eq:a5}
\end{equation}
is the mean brightness temperature of redshifted 21 cm radiation which
depends only on $z$ and  the background cosmological parameters ,
and 
\begin{equation}
  \d(\r,z)= \frac{\rho_{\HI}}{\bar{\rho}_{\H}} \left(1-\frac{\Tc}{\Ts}
  \right)  \left [1-\frac{(1+z)}{  H(z)}\frac{\partial v}{\partial
      r}\right]    
  \label{eq:a6}
\end{equation} 
is the  ``21  cm radiation  efficiency in redshift space''. The term 
$\d(\r,z)$ should be evaluated at $\r$, the position in real
space corresponding to $\z$  - calculated using eq. (\ref{eq:a1}))
without the effect of peculiar velocities, at the epoch corresponding
to the redshift $z$.
Here the term  $\rho_{\HI}/ \bar{\rho}_{\H}$ is the ratio 
of the neutral  hydrogen density to the mean hydrogen
density. Assuming that the hydrogen traces the dark matter and that
the neutral fraction varies from place to place we have  
$\rho_{HI}(\r,z)=\bar{\rho}_H(z)\, \bar{x}_{HI}(z)
[1+\delta_x(\r,z)][1+\delta(\r,z)]$.  
Here  $\delta_x $ and $\delta$  refer  to  fluctuations in
the neutral fraction and the  dark matter respectively, and
$\bar{x}_{HI}$ is the mean neutral fraction. The spin temperature
$\Ts$ also can vary from place to place i .e $\Ts(\x,z) =
\bar{\Ts}(z)[1+\delta_s(\x, \,z)] $. The last term in eq.
(\ref{eq:a6})  incorporates the effects of linear peculiar velocities
where $v(\r,z)$ is the line 
of sight component of the peculiar velocity and $\partial v/\partial r$ is
its radial derivative. We then have, at linear order,  the 
fluctuating part of $\d(\r,z)$  to be 
\begin{eqnarray}
  \d(\r,z)& = &\bar{x}_{HI}(z) \left\{  \left(1-\frac{\Tc}{\bar{\Ts}}\right)
  \left[\, \delta(\r,z) + \delta_x(\r,z) \right. \right.  \nonumber
  \\  & - & \left.\left.  \frac{(1+z)}{H(z)}\frac{\partial  
  v(\r,z)}{\partial r} \right]  +
  \frac{\Tc}{\bar{\Ts}}\,\delta_s(\r,z)\, \right\}
  \label{eq:a7}
\end{eqnarray} 

This is  most conveniently expressed in Fourier space as

\begin{equation}
  \d(\x,z)=\int \frac{d^3 k}{(2
    \pi)^3}\,\dt(\k,z) \, e^{-i \   \k \cdot \x} 
  \label{eq:a8}
\end{equation}
where
\begin{eqnarray}
  \dt(\k ,z)&=&\bar{x}_{HI}(z){\bigg \{}
  (1-\frac{\Tc}{\Ts}) \left[\Delta(\k,z)+\Delta_x(\k,z) \right.
\nonumber \\ &+& \left.  f \mu^2 \Delta(\k,z) \right]  + \frac{\Tc}{\Ts}
 \Delta_s(\k\,z)\,\,{\Bigg \}} 
\end{eqnarray}  
Here $\mu=\n \cdot \k/k$ is the cosine of the angle between the
Fourier mode $\k$  and the line of sight $\n$ towards the center of
the part of the sky under observation, and $f=dlnD/dlna$ where $D(a)$ is
the growing mode of density perturbations \citep{peebles} and $a$ is
the scale factor. The term
involving $f \, \mu^2$ arises from the peculiar velocities. 

The HI  correlation function $\xi_{\HI}({\bf \Delta r},z)=\langle \d(\r,z)
  \d({\bf r+\Delta r},z) \rangle$ and the HI power spectrum
  $P_{\HI}(\k,z)$ defined as $\langle \dt(\k,z)
  {\dt}^{*}(\k^{'},z) \rangle=(2 \pi)^3 \delta^3(\k-\k^{\prime})
  P_{\HI}(\k,z)$  are related as 
\begin{equation}
 \xi^s_{HI}({\bf \Delta r},z)=\int \frac{d^3 k}{(2\pi)^3} e^{ - i\,\bf 
  k\cdot\,{\bf \Delta r}} P_{HI}({\bf k},z).
      \label{eq:a10}
\end{equation}
where the HI power spectrum is 
\begin{eqnarray}
    P_{HI}({\bf{k}} ,z)& = &{\bar{x}}^2_{HI}(z)
\,{\bigg \{}\left(1-\frac{\Tc}{\bTs} \right)^2\, [(1+ 
      f\,{\mu}^2)^2\,P_{\delta-\delta}(k ,z)\nonumber\\ &+& 2(1+\
      f\,{\mu}^2)\,P_{\delta-x}(k ,z)+P_{x-x}(k,
      z) ]\, \nonumber \\ &+& 2
\left(1-\frac{\Tc}{\bTs} \right)\,\frac{\Tc}{\bTs}
[(1+f\,{\mu}^2)\,P_{\delta-s}(,z)\nonumber \\ &+& P_{x-s}({\bf{k}} ,z)]
+\left(\frac{\Tc}{\bTs} \right)^2\,P_{s-s}(k ,z)\,\,{\Bigg \}}
\label{eq:a11}
\end{eqnarray}
Here $P_{\delta-\delta}(k,z)$, $P_{x-x}(k,z)$ and $P_{s-s}(k,z)$ are
the power spectra of the three different sources which contribute to
the HI  fluctuations, the
dark matter, neutral fraction and spin temperature fluctuations
respectively. The cross power spectra $ P_{\delta-x}(k ,z)$,
$P_{\delta-s}(k ,z)$ and  $P_{x-s}(k ,z)$ quantify
cross-correlations, if any,  between these different sources. 
The presence of peculiar velocities makes the HI power spectrum
$P_{\HI}(\k,z)$ anisotropic through the terms involving $f \mu^2$
which appear in eq. (\ref{eq:a11}). It should be noted that
$f \mu^2$ appears only with the sources of HI fluctuation that  are
correlated with the dark matter fluctuations, and the contribution to
the HI power spectrum from components not correlated with the
dark matter are isotropic. 

The anisotropy of $P_{\HI}(\k,z)$ is reflected in $\xi_{\HI}({\bf
  \Delta r},z)$ which depends on how ${\bf \Delta r}$ is aligned with
  respect to $\n$. Further, $\xi_{\HI}({\bf \Delta r},z)$ depends only
  on the even powers of $ \n \cdot {\bf \Delta r}/\Delta r$. Shifting
  our focus back to $\w(\thl,\thp)$ we have  
\begin{equation}
 \w(\thl,\thp)=\bar{T}^2(z) \,. \xi^s_{\HI}({\bf \Delta r},z) 
\end{equation}

The correlation in the HI brightness temperature fluctuations,  $
  \w(\thl,\thp)$ is anisotropic due to two reasons, first because the
  mapping from $(\thl,\thp)$ to ${\bf \Delta r}$ is anisotropic (the
  AP effect) and second because $\xi_{\HI}$ is intrinsically
  anisotropic due to the effect of peculiar velocities.  Here we
  investigate whether it will be possible to observationally
  distinguish these  two distinct sources of anisotropy. Both
  these effects depend on the  background cosmological model,
  and we investigate to what extent observations of the anisotropy
  can be used  to determine the cosmological parameters. Finally, the 
anisotropy in the HI power spectrum depends on the
 properties of the different sources which contribute to HI
  fluctuations.  We investigate to what extent observations of
  anisotropies in the HI can disentangle the contributions from the
  different sources. 

\section{Models}
\subsection{Background Cosmology}
We have considered a few representative cosmological models,  all of
which are spatially flat and have two components namely dark matter
and dark energy. The possibility that the dark energy equation of
state evolves with time has, of late, received a considerable amount
of attention (eg.  \citealt{peebles1} , \citealt{ries} ,
 \citealt{perl} , \citealt{sahni} , \citealt{sahni1} ,
 \citealt{saini1}, \citealt{carroll}, \citealt{hut}, \citealt{perl1},
 \citealt{linder}, \citealt{alam} ) . Here we consider a possible
parameterization of the variable dark energy  (VDE) model  introduced 
by \citet{wang}.  
The Hubble parameter is of the form 
\begin{equation}
  H(z)= H_0\,[\Omega_{m 0}\,(1+z)^{3} + (1-\Omega_{m 0})
  f_{DE}(z)]^{1/2}.  
  \label{eq:a2}
\end{equation} 
where $H_0$, $\Omega_{m0}$, and $1-\Omega_{m0}$ are the present
values of the Hubble parameter, the dark matter density parameter and
the dark energy density parameter respectively, and 
 $f_{DE}(z) 
    =\rho_{DE}(z)/\rho_{DE}(0) = f_{\infty} + (1 -
    f_{\infty})\,\,(1+z)^{3(1+w_i)} $
is  the dimensionless dark energy density function.  This VDE model has
two extra parameters $f_{\infty}$ and $w_i$  
compared to  the usual LCDM model, and the parameters
are restricted to the range $w_i\ge -2$ and $f_{\infty} \ge 0$. 
The VDE model   reduces to to LCDM model if either
$f_{\infty}=1$ or $w_i=-1$. In addition to the SCDM model
($\Omega_{m0}=1$) and the LCDM model with $\Omega_{m0}=0.3$, we
have considered the VDE model with $\Omega_{m0}=0.4$ and
$(f_{\infty},w_i)=(1.2,-1.8)$. In the VDE models which we
have considered,  $f_{DE}(z)$ increases from $f_{DE}(0)=1$ and
saturates at $f_{\infty}=1.2$ around $z \sim 3$. The value of $f_{DE}$
reaches the saturation value at a lower redshift  is $w_i$ is decreased. 
Figures \ref{fig:a1} and \ref{fig:a2} show the behavior of $p(z)$ and
$f(z)$ respectively for the models considered here. The points
to note are (1.) Though in all models $p(z)$ falls rapidly up to $z
\sim 5$ beyond which it decreases gradually, the values of $p(z)$ are
different in each of the models, and  (2.) The values of $f(z)$ differ
from model to model only for $z < 4$, and $f(z) \sim 1$ at larger
redshifts where the universe is dark matter dominated. 
\subsection{The HI distribution}
We focus our attention on two different epochs which we discuss below.
\subsubsection{Reionization}
We adopt  a simple model for the hydrogen distribution where there are
spherical ionized bubbles of comoving radius $R$ and the region
outside the bubbles is completely neutral. The total hydrogen density
is assumed to trace the dark matter distribution. The reionization is
believed to have been caused by the first luminous objects which are
expected to be highly clustered.  We incorporate this by assuming that
the centers of the ionized bubbles are biased with respect to the dark
matter distribution with a bias $b_c$.  Further, we assume that the
hydrogen has been heated before reionization {\it ie.} $\Ts \gg \Tc$
and consequently the HI will be observed in emission. 
The fluctuations in the neutral fraction $(\delta_x)$ has two parts,
one which is correlated with the dark matter fluctuations and another,
arising from the discrete nature of the ionized regions, which is
uncorrelated with the dark matter distribution. The spin temperature
fluctuations $(\delta_s)$ do not contribute when $\Ts \gg \Tc$.  
The power spectrum for the
resulting HI distribution is \citep{bharad6}
\begin{eqnarray}
P_{\HI}(\k,z)&=&\left[\bar{x}_{\HI} (1+ f \mu^2) - b_c f_V W(k
  R) \right]^2  P_{\delta-\delta}(k,z)  \nonumber \\
&+& \frac{f^2_V W^2(k R) }{\bar{n}_{\HI}} 
\label{eq:c3}
 \end{eqnarray}
 where $W(y)=(3/y^3)[\sin(y) - y \cos(y)]$ is the spherical
top hat window function, 
$f_V$ is the fraction of volume which is ionized,
$\bar{n}_{\HI}$ is the comoving number density of the spheres and $f_V=(1-
\bar{x}_{\HI})=\frac{4}{3} \pi R^3 \bar{n}_{\HI}$. 

of spherical HII region
The first term which contains $P_{\delta-\delta}(k)$ arises  from the
clustering of 
the hydrogen and the clustering of the centers of the ionized
spheres. The second term which has $1/\bar{n}_{\HI}$  is the Poisson
noise  due to 
the discrete nature of the ionized regions. The latter is not
correlated with the dark matter.

This  model has a limitation that  the HI density is negative in
a fraction $ \sim f_V^2/2$  of the total volume where ionized spheres
overlap. The possibility of the spheres overlapping  increases if  
they are highly clustered. This restricts the range of $f_V$   and 
 $b_c$ where this model is meaningful.

\subsubsection{Post-reionization}
At low redshifts the bulk of the HI is in the high column density
clouds which produce the  damped Lyman-$\alpha$ absorption lines
observed  in  quasar spectra (\citealt{peroux} , \citealt{lombardi} ,
\citealt{lanzetta}). These observations currently indicate
$\Omega_{gas}(z)$ , the comoving 
density of neutral gas expressed as a fraction of the present critical 
density, to be  nearly constant at a value   $\Omega_{gas}(z) \sim
10^{-3}$ for   $z \ge 1$ \citep{peroux}. The damped Lyman-$alpha$
clouds are believed to be associated with galaxies which represent 
highly non-linear overdensities. 
  It is now generally accepted
from the study of the large scale structures in redshift surveys
and  N-body simulations that the galaxies (or nonlinear structures)
are a biased tracer of the underlying dark matter distribution
(\citealt{kiser} ;\citealt{dekel};\citealt{mo} ;\citealt{taru} and 
\citealt{yosi}). On the large scales of interest here it is reasonable
to assume that these HI clouds trace the dark matter with a constant,
linear bias $b$. 

Converting $\Omega_{gas}$ to the mean neutral fraction
$\bar{x}_{\HI}=\bar{\rho}_{\HI}/\bar{\rho}{\H}=\Omega_{gas}/\Omega_b$
gives us $ \bar{x}_{\HI}=50 \Omega_{gas}  h^2 (0.02/\Omega_b h^2)$ or
$\bar{x}_{\HI}=2.45 \times 10^{-2}$. We also assume $\Ts \gg \Tc$ and
hence we see the HI in emission.   Using these we have 
\begin{equation}
P_{\HI}(\k ,z)=\bar{x}^2_{\HI}\left( 1+ \beta  \mu^2 \right)^2
\,P_{\delta-\delta}(k ,z) 
\label{eq:a26}
\end{equation}
where $\beta=f/b$. 
The fact that the neutral hydrogen is in discrete clouds  makes a
contribution which we do not include here.  Another important effect
not included here is that the fluctuations become non-linear at low
$z$. Both these effects have been studied using simulations
\citep{bharad4}. 

\section{Results}
We quantify the anisotropies in the HI clustering pattern by
decomposing $\w(\thl,\thp)$ into different spherical harmonics 
  \begin{equation}
    \w_l{(\theta)} = \frac{2\,l+1}{2}\int^1_{-1}
\w{(\theta,\,\tilde{\mu})}\, \, P_l(\tilde{\mu})\,d\tilde{\mu}
	\label{eq:a19}
  \end{equation}
  where $P_l(\mut)$ are the Legendre polynomials and we have used
 $\w(\theta,\mut)= \w(\thl,\thp)$ where $\theta=\sqrt{\thl^2+\thp^2}$ and
 $\mut=\thl/\theta$. The fact that $P_{\HI}(\k,z)$  depends only on
 even powers of $\mu$ ensures that all the odd harmonics will be
 zero.  In addition to the monopole $\w_0(\theta)$, we have 
calculated  the quadrupole $\w_2(\theta)$ and the hexadecapole
 $\w_4(\theta)$ and we use  the ratios  $\w_2/\w_0$ and $\w_4/\w_0$ to
 quantify the anisotropies. 

\subsection{Reionization}

\begin{figure}
 \rotatebox{-90.0}{\scalebox{0.35}{\includegraphics{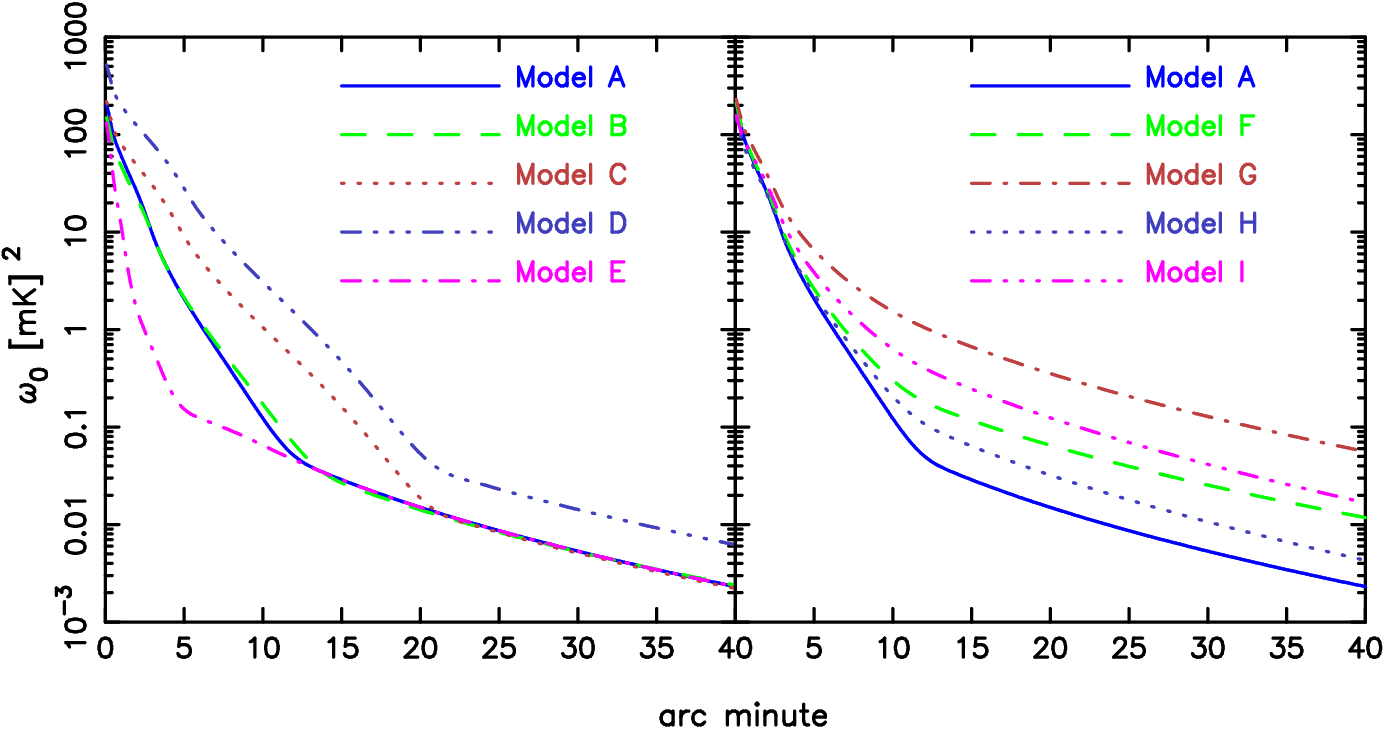}}}
  \caption{For the different models shown in Table~1, this shows the
    monopole  of the signal expected at the epoch of reionization. 
Here $1^{'}$ corresponds to $1.9 h^{-1}{\rm Mpc}$ in the LCDM.}
\label{fig:3}
\end{figure}

\begin{figure}
 \rotatebox{-90.0}{\scalebox{0.35}{\includegraphics{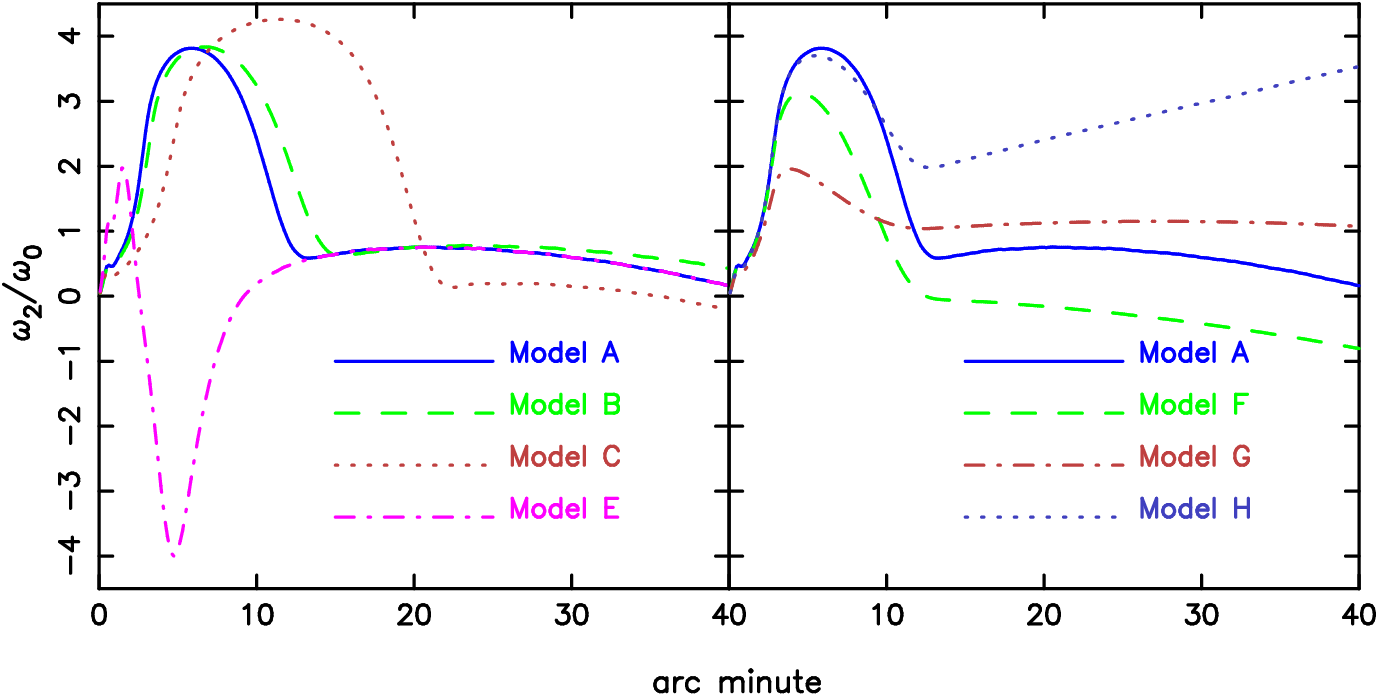}}}
  \caption{For the different models shown in Table~1, this shows the
quadrupole to  monopole  ratio  of the signal expected at the epoch of
reionization. Here $1^{'}$ corresponds to $1.9 h^{-1}{\rm Mpc}$ in the
LCDM. } 
\label{fig:4}
\end{figure}

\begin{figure}
 \rotatebox{-90.0}{\scalebox{0.35}{\includegraphics{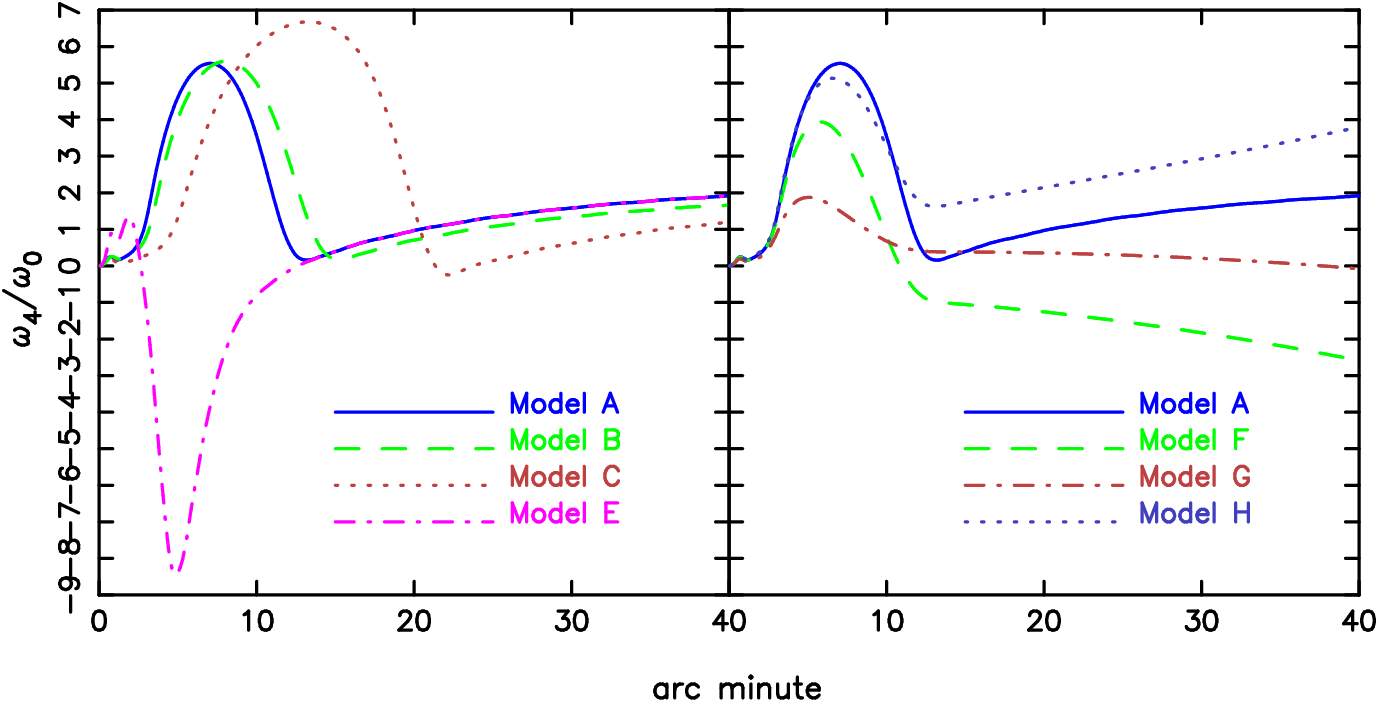}}}
  \caption{For the different models shown in Table~1, this shows the
hexadecapole  to  monopole  ratio  of the signal expected at the epoch of
reionization. Here $1^{'}$ corresponds to $1.9 h^{-1}{\rm Mpc}$ in the
LCDM. }  
\label{fig:5}
\end{figure}

We have restricted our analysis to $z=10$ which corresponds to $129
\, {\rm MHz}$ and  assume that half the hydrogen is ionized at this 
redshift {\it ie.} $f_V=0.5$  \citep{zald}.
 Recent investigations indicate that at this redshift
the comoving size of the ionized bubbles  will be of
the order of a few  ${\rm Mpc}$  \citep{furlanetto1}. Further,
\citet{furlanetto1} also show that the bias is expected to have a low
value (near unity)  for large bubble size  $R$. 
We  consider the LCDM model with $R=3 \, h^{-1} {\rm Mpc}$ and
$b_c=1.5$ as the fiducial model (Model A) for which Figures
\ref{fig:3}, \ref{fig:4} 
and \ref{fig:5} show $\w_0$, $\w_2/\w_0$ and $\w_4/\w_2$ respectively.
To study how the signal depends on various factors, we have varied the  
background cosmological model and  the parameters $R$ and $b_c$. 
Further, we have also considered the signal without  the effect of the 
peculiar velocities and the Poisson term in order to explicitly
isolate the contribution from these terms. Table~1 shows the various
combinations for which we have calculated the expected signal. 

One of the salient  features  of the monopole $\w_0$ (Figure
\ref{fig:3}) is that the signal is dominated by the Poisson
fluctuations from the discrete  ionized
bubbles  on small scales   whereas it traces the dark matter on
large scales.  The angular scale where  the transition from 
Poisson fluctuations to  dark matter occurs  depends on the background
cosmological model.  Further, the Poisson contribution increases if
the bubbles are larger (increasing $R$) because the number density of
bubbles falls.  The dependence on the background cosmological model 
 can be attributed to the
fact that the comoving distance corresponding to $z=10$  differs by 
around $10 \%$ in the LCDM and VDE models whereby  the same ionized
bubbles  
correspond to  different angles in the two models. 
It should be noted that though the overall amplitude of $\w_0$ also
changes with the background cosmological model, it would not be
possible to 
distinguish this from a change in the neutral fraction which would
have the same effect. Increasing $b_c$ increases the overall-amplitude
of the signal at large scales leaving it unchanged at very small
scales. The effect of peculiar velocities, we find, depends on the
value of $b_c$, and it increases the signal when $b_c=0$ whereas  it
reduces the signal when $b_c=1.5$. An interesting situation occurs
when $b_c=1$  where the coefficient of $P_{\delta-\delta}(k)$ in
eq. (\ref{eq:c3}) nearly cancels out and the signal is extremely small
on large scales. 
\begin{table}

\caption{This shows the different combinations for which we have
  calculated the signal expected at the epoch of reionization. Here
  ${\rm R }$ is in $ h^{-1} {\rm   Mpc}$. Further,  $\rm P.V $ and
  $\rm P.F$ 
  respectively   indicate whether  the peculiar velocities and  the
  Poisson fluctuations have been included. }

\label{table1}
\begin{tabular}{@{}lccccccc}
Model &  Background cosmology  &  $\rm R$ & $b_c$  & $\rm P.V $  &  $\rm P.F$ &\\
$\rm A$    &      $\rm LCDM$  &  3 &  1.5     & $\surd$   &  $\surd$\\
$\rm B$    &      $\rm VDE$   &  3 &  1.5     & $\surd$   &  $\surd$\\
$\rm C$    &      $\rm LCDM$  &  5 &  1.5     & $\surd$   &  $\surd$\\
$\rm D$    &      $\rm VDE$   &  5 &  1.5     & $\surd$   &  $\surd$\\
$\rm E$    &      $\rm LCDM$  &  3 &  1.5     & $\surd$   &  $\times$\\
$\rm F$    &      $\rm LCDM$  &  3 &  1.0     & $\surd$   &  $\surd$\\
$\rm G$    &      $\rm LCDM$  &  3 &  0.0     & $\surd$   &  $\surd$\\
$\rm H$    &      $\rm LCDM$  &  3 &  1.5     & $\times$  &  $\surd$\\
$\rm I$    &      $\rm LCDM$  &  3 &  0.0     & $\times$  &  $\surd$\\

\end{tabular}
\end{table}

Considering next  the anisotropy (Figure \ref{fig:4} and
\ref{fig:5}) we find that the dominant feature is a bump at small
scales caused by the 
Poisson fluctuations from the discrete  ionized
bubbles. The location of the bump is sensitive to the size of the
bubbles.  The nature of the anisotropy is significantly altered if
the Poisson fluctuations are not taken into account.  The bump in the
anisotropy is a consequence of the AP effect, as can be inferred from
the fact that it is not changed much if the peculiar velocities are
not taken into account. Further, this feature is not very sensitive to
the background cosmological model. The bias $b_c$ of the ionized
bubbles changes the amplitude of the bump. The large scale 
anisotropy is a combination of both the peculiar velocities and the AP
effect which make opposite  contributions to the anisotropy. The large
scale anisotropy is nearly constant at a value which depends on both
$b_c$ and $R$. Finally we note the fact that the signal is highly
anisotropic with $\w_4/\w_0 > \w_2/\w_0 >1$, and it is possible that
some of the higher angular moments not considered here have large
values compared to $\w_0$.   

\subsection{Post-reionization}
\begin{figure}
  \rotatebox{-90.0} {\scalebox{0.5}{\includegraphics{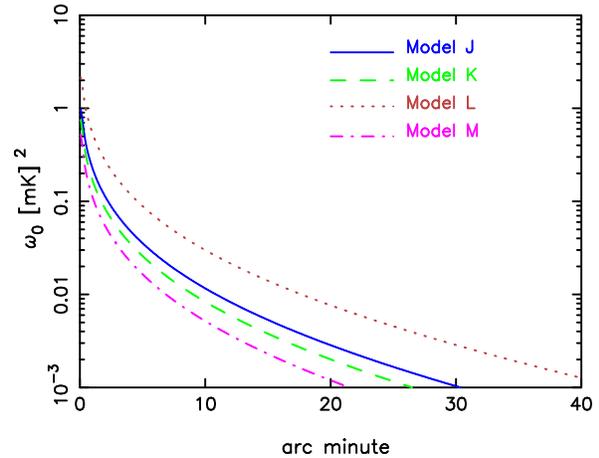}}}
  \caption{For the different models shown in Table~2, This shows the
    monopole  for signal expected at  $325 \, {\rm MHz}$.
Here $1^{'}$ corresponds to $1.4 h^{-1}{\rm Mpc}$ in the LCDM. } 
  \label{fig:6}
\end{figure}

\begin{figure}
  \rotatebox{-90.0} {\scalebox{0.5}{\includegraphics{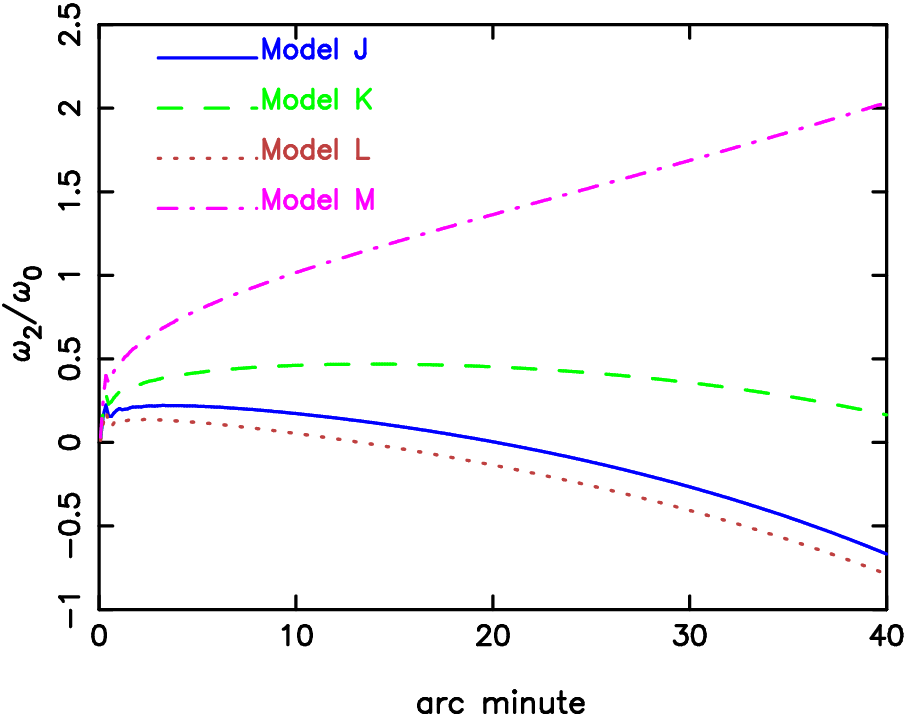}}}
  \caption{For the different models shown in Table~2, this shows the
quadrupole to  monopole  ratio  of the signal expected at  $325 \, {\rm
  MHz}$. Here $1^{'}$ corresponds to $1.4 h^{-1}{\rm Mpc}$ in the LCDM.
}
  \label{fig:7}
\end{figure}

\begin{figure}
  \rotatebox{-90.0}{\scalebox{0.5}{\includegraphics{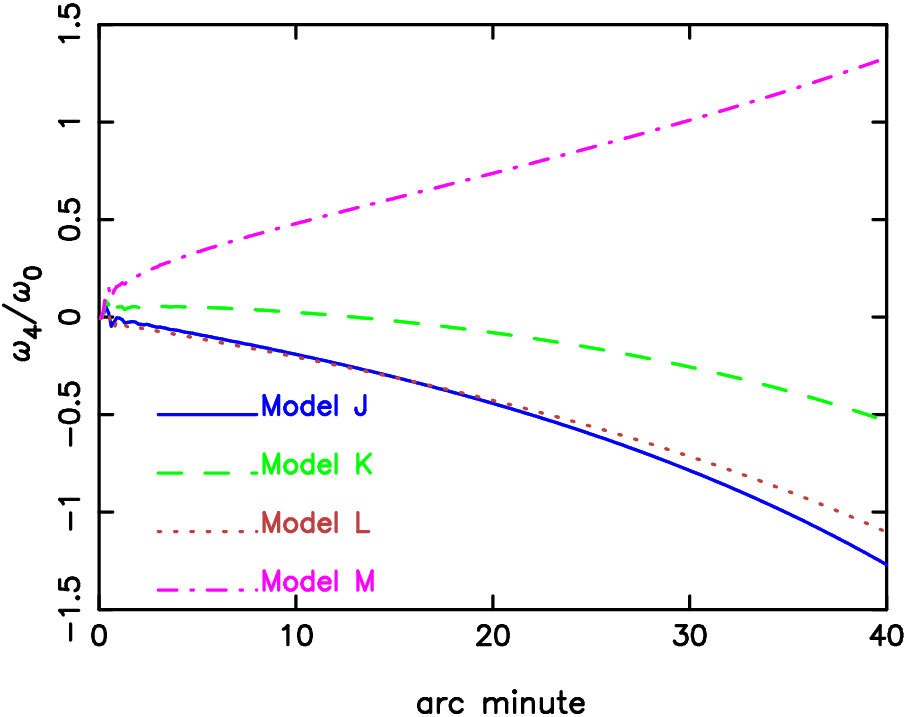}}}
  \caption{For the different models shown in Table~2, this shows the
hexadecapole  to  monopole  ratio  of the signal expected at  $325 \, 
{\rm MHz}$ . Here $1^{'}$ corresponds to $1.4 h^{-1}{\rm Mpc}$ in the
LCDM. } 
  \label{fig:8}
\end{figure}

We restrict our analysis here to $z=3.37$ which corresponds to
frequency $325 \, {\rm MHz}$ for the HI radiation. In addition to the
possibility of different background cosmological models, there is only
one free parameter $b$ namely the bias  of the HI relative to the dark
matter. We  consider the LCDM model with $b=1$ as the fiducial model
(Model J) for which Figures \ref{fig:6}, \ref{fig:7} and \ref{fig:8}
shows $\w_0$, $\w_2/\w_0$ and $\w_4/\w_0$ respectively. The other
models which we have considered are listed in Table~2.   

The monopole $\w_0$ (Figure \ref{fig:6}) traces the dark matter
distribution. Though the amplitude varies in the different models not
much significance can be attached to this as such an effect  can also
arise from variations in the neutral fraction which is quite
uncertain.   The anisotropy in the clustering
pattern is a combination of the peculiar velocities and the AP
effect. This is deduced  from the large change in the 
anisotropy when  the peculiar velocity contribution  is dropped. 
Both the effects  are sensitive to the  the background
cosmological model at this redshift (Figures \ref{fig:a1} and
\ref{fig:a2}).  The point to note is that  changing the bias $b$ has
an effect which is very similar to that of changing the
background cosmological model, and it will be hard to use observations
of the anisotropies to individually constrain any one of them. 
  
\begin{table}

\caption{ This shows the different combinations for which we have
  calculated the signal expected at post-reionization. Here
  ${\rm b}$ is bias. Further,  $\rm P.V $  indicates the peculiar
  velocities of HI.}

\label {table2}
\begin{tabular}{@{}lccccccc}
Model &  Background cosmology  & $b$    & $\rm P.V $  & \\
$\rm J$    &      $\rm LCDM$   &  1     & $\surd$     &  \\
$\rm K$    &      $\rm LCDM$   &  1.8   & $\surd$     & \\
$\rm L$    &      $\rm VDE$    &  1     & $\surd$     &  \\
$\rm M$    &      $\rm LCDM$   &  1     & $\times$    & \\

\end{tabular}
\end{table}

\section{Discussion and conclusions.}
The AP effect and the peculiar velocities both introduce anisotropies
in the redshift space HI clustering pattern. These two sources of
anisotropy are  parameterized by two different functions $p(z)$
(Figure \ref{fig:a1}) and $f(z)$ (Figure \ref{fig:a2}) respectively,
both of which depend on the background cosmological model. We have
focused on  two different background models. One of which has
$\Omega_{m0}=0.3$ and $\Omega_{\Lambda0}=0.7$ and another VDE model has $\Omega_{m0}=0.4$ and $\Omega_{\Lambda0}=0.6$.  The two models differ
in that in one of the models the dark energy density remains a
constant (LCDM) whereas in the other model (VDE) it increases with
redshift and saturates at a value thrice  the present value at $z \sim
3$. We find that $f(z)$ varies with the cosmological model only at low
redshifts and it saturates at $f(z)=1$ at high redshifts where the
universe is dark matter dominated in most cosmological models. Thus
the anisotropies introduced by the peculiar velocities will be
sensitive to the background cosmological model only at low redshifts,
and 
it cannot  tell us  much about the background cosmological model at
high redshifts.   The function $p(z)$ is sensitive to the background
cosmological models at nearly all redshifts $z > 0.1$, and the
anisotropies introduced by the AP effect hold the possibility of
allowing us to probe the background cosmology at high redshifts.    

We have considered simple models for the HI distribution at the epoch
of reionization and in the post-reionization era.  These models have a
few parameters which quantify our ignorance  about the HI distribution
during these epochs. For the reionization era we have assumed that the
total hydrogen content traces the dark matter, and the reionization
proceeds through  spherical bubbles of ionized gas of  comoving radius
$R$.  Further, the centers of these bubbles could be biased with
respect to the underlying dark matter. The HI in the post reionization
era is assumed to be in high column density clouds which could be
biased with respect to the underlying dark matter. 
For both the epochs we find that the anisotropies in the HI
clustering pattern are sensitive to both, the background cosmological
model and the free parameters determining the HI distribution. Given the
fact that very little is known a-priori about the high redshift
HI distribution, we conclude that it will not be possible to use 
observations of the anisotropies alone to uniquely constrain the
background cosmological model. It may be noted that our findings
contradict earlier claims \citep{adi} where it was proposed that the
anisotropies in the epoch of reionization HI signal could be used to
constrain  the background cosmological model.  
 
We find that the anisotropy in the HI clustering is a combination of
the contributions from both the AP effect and the peculiar velocities,
and it cannot be attributed to anyone of them alone. Further, the
anisotropy pattern at the epoch of reionization is very sensitive to
the reionization model.  In our simple model where the ionized
bubbles are all of the same radius, we find a bump in the anisotropy
pattern  at the scales corresponding to the size of the 
bubbles.   The bump will possibly be smeared  in a more realistic
situation  where the bubbles will have a spread of sizes, but we may
still expect a prominent feature at the scale corresponding to the
characteristic bubble size. The anisotropy at large angular scales, we
find, is  sensitive to the bias of the centers of the ionized bubbles.   
Our analysis leads us to the conclusion that  observations of the
anisotropy in the HI clustering at the epoch of reionization will be a
very powerful  probe of the size, shape and distribution of the 
ionized regions the uncertainties in these being much more than the
uncertainties in the background cosmological model. In the
post-reionization era, the changes  in the value of the bias
parameter have a similar effect as changing the background
cosmological model. These observations can be used to constrain the
cosmological model if the bias parameter can be determined by
independent means like the bi-spectrum (\citealt{verde};
\citealt{scoc}).

Finally we note that the temperature two-point correlation function
considered here may not be the optimal statistical tool for detecting
and quantifying the high redshift HI signal. The angular power
spectrum \citep{zald} and the visibility correlations
(\citealt{bharad2}, \citealt{morales} and \citealt{bharad6}) have been
proposed as optimal statistics for this purpose. In this paper we have
chosen to study   the temperature  
two-point correlation because of its similarity to the galaxy
two-point correlation function where the 
anisotropy introduced by redshift space distortions is well understood
and has received much attention in the literature.   Though both the
AP effect and the peculiar velocities are both included in
\citet{bharad2}  and \citet{bharad6},  how to make the best use of
these observations to constrain background cosmological models and
models for the HI distribution still remains an open issue. 
\section*{Acknowledgments}
 SB would like to acknowledge BRNS, DAE, Govt. of India,for
 financial support through sanction No. 2002/37/25/BRNS. SSA would like
 to thank  Kanan Kumar Datta for useful discussions.    
SSA and BP would like to acknowledge the CSIR, Govt. of India for financial
 support through  a senior research fellowship.

\end{document}